\documentclass[journal]{IEEEtran}
\usepackage{bm}
\usepackage{cite}
\usepackage{amsmath,amssymb,amsfonts}
\usepackage{algorithmic}
\usepackage{graphicx}
\usepackage{textcomp}
\usepackage{xcolor}
\usepackage{subfigure}

\def\BibTeX{\rm B\kern-.05em{\sc i\kern-.025em b}\kern-.08emT\kern-.1667em\lower.7ex\hbox{E}\kern-.125emX}
\allowdisplaybreaks[4]

\newtheorem{lemma}{Lemma}

\begin{document}

\title{NOMA Enabled Multi-Access Edge Computing: A Joint MU-MIMO Precoding and Computation Offloading Design}

\author{Deyou Zhang, Meng Wang, Shuo Shi, and Ming Xiao \vspace{-2em}

\thanks{D. Zhang and M. Xiao are with the School of Electrical Engineering and Computer Science, KTH Royal Institute of Technology, 10044 Stockholm, Sweden (email: \{deyou, mingx\}@kth.se). M. Wang and S. Shi are with the School of Electrical and Information Engineering, Harbin Institute of Technology, 150001 Harbin, China (email:\{wangmeng\_1996, crcss\}@hit.edu.cn).}
}

\maketitle

\begin{abstract}
    This letter investigates computation offloading and transmit precoding co-design for multi-access edge computing (MEC), where multiple MEC users (MUs) equipped with multiple antennas access the MEC server in a non-orthogonal multiple access manner. We aim to minimize the total energy consumption of all MUs while satisfying the latency constraints by jointly optimizing the computational frequency, offloading ratio, and precoding matrix of each MU. For tractability, we first decompose the original problem into three subproblems and then solve these subproblems iteratively until convergence. Simulation results validate the convergence of the proposed method and demonstrate its superiority over baseline algorithms.
\end{abstract}

\begin{IEEEkeywords}
    Multi-access edge computing, non-orthogonal multiple access, computation offloading, precoding design.
\end{IEEEkeywords}

\IEEEpeerreviewmaketitle

\section{Introduction}
\IEEEPARstart{W}{ith} the continuous growth of mobile services, portable terminals (e.g., smartphones or tablet computers) are running more and more computation-intensive and latency-critical applications, which brings new challenges to those terminals' batteries and central processing units (CPUs) \cite{survey2}. To cope with this issue, multi-access edge computing (MEC) has been proposed. As reported in \cite{survey3}, by deploying MEC servers to the network edge, terminals' huge computational burden and energy consumption can be greatly reduced.

Computation offloading and resource allocation (including both communication and computation resources) are the main challenges in MEC, and many insightful works have been arisen in this field such as \cite{Xidian, survey7, massivemimo}. Specifically, the authors in \cite{Xidian} focused on a single-user multiple-server scenario and proposed to jointly optimize the offloading ratio, the user's computational speed and transmit power to achieve two purposes: user energy consumption minimization or task execution latency minimization. It is worth mentioning that both the user and servers are assumed to be equipped with a single antenna in \cite{Xidian}. The authors in \cite {battery} further considered a MIMO multicell MEC system where multiple users ask for computation offloading to a common server. In that paper, the authors aimed to jointly optimize the users' transmit precoding matrices and the server's CPU cycles/second assigned to each user to minimize the total energy consumption of all users. Moreover, a binary offloading problem was considered in \cite{mumimo}, where the authors proposed to jointly optimize the offloading decision making, users' precoding matrices, and the server's CPU cycles/second assigned to each user to minimize a weighted sum of energy consumption and time delay of all users.

In parallel with the development of MEC, non-orthogonal multiple access (NOMA) has been recognized as a promising technology to achieve high spectral efficiency \cite{mimonoma}, such that more and more works have investigated the integration of NOMA and MEC. Specifically, literature \cite{decoding1} proposed an edge computing aware NOMA technique with the aim of employing uplink NOMA to reduce the users' uplink energy consumption. Similarly, a joint radio and computation resource allocation problem for NOMA-based MEC in heterogeneous networks was studied in \cite{decoding2}, where the authors aimed to minimize the total energy consumption of all MEC users considering the task execution latency constraint. Moreover, the authors in \cite{mecnoma2} integrated NOMA-based MEC into the Internet of Things (IoT), enabling the IoT devices to offload their delay-sensitive and computation-intensive tasks to the network edge. Although existing works have investigated the integration of NOMA and MEC, to the best of our knowledge, there is still no work considering the joint computation offloading and transmit precoding design for MIMO-NOMA based MEC systems. The most similar work in this field is \cite{multinoma}, but only the MEC server is assumed to be equipped with multiple antennas and each user still has a single antenna.

Motivated by the above discussions, the purpose of this letter is to investigate the computation offloading and transmit precoding co-design for MIMO-NOMA empowered MEC. More specifically, we consider partial task offloading and aim to jointly optimize the computational frequency, offloading ratio, and transmit precoding matrix of each MEC user (MU) to minimize their total energy consumption under the task execution latency constraints. To handle this intractable problem, we first decompose it into three subproblems, i.e., computational frequency optimization subproblem, offloading ratio optimization subproblem, and transmit precoding design subproblem, and then solve them iteratively until convergence. Simulation results confirm the convergence of the proposed method and demonstrate its superiority over baseline algorithms.

\textbf{Notations}: In this paper, the sets of complex-valued and real-valued matrices with dimension $M \times N$ are respectively denoted by ${\cal C}^{M \times N}$ and ${\cal R}^{M \times N}$. Upper case boldface letters (e.g., $\mathbf A$) denote matrices and lower case boldface letters (e.g., $\mathbf a$) indicate vectors. $\mathbf I$ indicates the identity matrix. $\mathbb{E}[\cdot]$ is the expectation operator. The conjugate transpose, trace, and determinant of a matrix (e.g., $\mathbf A$) are respectively denoted by ${\mathbf A}^{\textsf H}$, $\text{Tr}(\mathbf A)$, and $|\mathbf A|$.

\begin{figure}[t]
	\centering
	\includegraphics[scale=0.3]{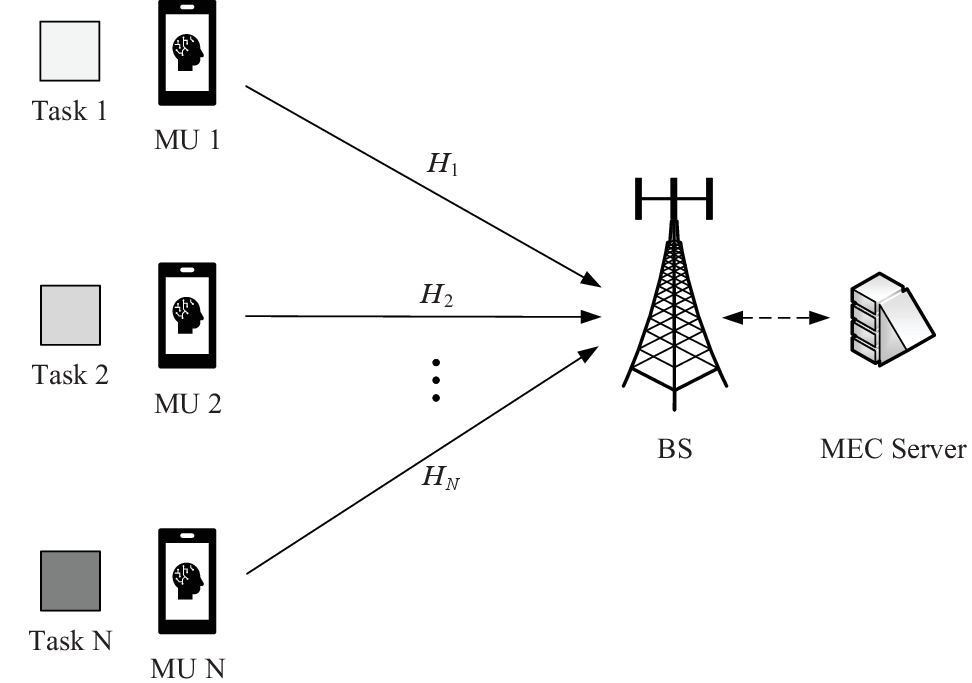}
	\caption{The considered MIMO-NOMA enabled MEC system.} \label{Fig:1}
\end{figure}

\section{System Model and Problem Formulation}
As depicted in Fig. \ref{Fig:1}, we consider an MEC system which consists of one base station (BS) and $N$ MUs. The BS that is embedded with an MEC server has $N_r$ antennas, and each MU has $N_t$ antennas, where $N_t < N_r$. Each MU has a data-partitioned-oriented task to deal with, and the task can be processed either locally or offloaded to the MEC server in a partial offloading manner \cite{p_offloading}. Following the existing literatures, we characterize the task of MU $k$ as $(L_k, C_k, T_k)$, $\forall k \in {\cal N} \triangleq \{1, \cdots, N\}$, where $L_k$, $C_k$, and $T_k$ respectively denote the data size (in bits), the number of CPU cycles required for processing a unit bit, and the maximum tolerable delay (in seconds). In this paper, we leverage the MIMO-NOMA technique to improve the offloading efficiency. That is, the MUs use the same time-frequency resources to transmit their data streams to the BS, which then performs a minimum mean square error equalizer with successive interference cancellation (MMSE-SIC) to decode the data streams of each MU. While there are a total number of $N!$ different decoding orders for the $N$ MUs, for convenience, we take the decoding order, $\text{MU}_1 \to \text{MU}_2 \cdots \to \text{MU}_N$, as an example to introduce the considered energy consumption minimization problem. Specifically, under the decoding order of $\text{MU}_1 \to \text{MU}_2 \cdots \to \text{MU}_N$, when the data streams of MU $k$ remain to be decoded, the residual signal at the BS can be expressed as
\begin{equation}
    \mathbf y_k = {\bf H}_k {\bf F}_k {\bf x}_k + {\bf Z}_k + {\bf n},
\end{equation}
where $\mathbf H_k \in {\cal C}^{N_r \times N_t}$, $\mathbf F_k \in {\cal C}^{N_t \times N_t}$, and $\mathbf x_k \in {\cal C}^{N_t \times 1}$ respectively denote the channel matrix, the precoding matrix, and the transmit symbol vector of MU $k$. As in \cite{mumimo}, we assume $\mathbf x_k \sim {\cal CN}(\mathbf 0, \mathbf I)$, such that the average transmit power of MU $k$ is given by ${\mathbb E}[\|\mathbf F_k \mathbf x_k\|^2] = \text{Tr}(\mathbf F_k \mathbf F_k^{\textsf H})$, $\forall k \in {\cal N}$. Moreover, we assume $\mathbb{E}[\mathbf x_k \mathbf x_i^{\textsf H}] = \mathbf 0$, $\forall i \ne k$, i.e., the data from different MUs are uncorrelated. The variable $\mathbf Z_k$ is the possible inter-user interference: $\mathbf Z_k = \sum\nolimits_{i = k + 1}^{N} {\mathbf H_i} \mathbf F_i \mathbf x_i$, $\forall k \in {\cal N}_{-1} \triangleq \{1, \cdots, N - 1\}$, and $\mathbf Z_N = \mathbf 0$. Last, $\mathbf n \sim {\cal CN}(\mathbf 0, \epsilon^2 \mathbf I)$ is the noise vector. By definition, $\epsilon^2 \triangleq N_0 B$, where $N_0$ and $B$ respectively denote the noise power spectral density and system bandwidth. According to \cite{mumimo}, the achievable uplink transmission rate of the $k$-th MU using the MMSE-SIC technique can be given by
\begin{equation}
    R_k = B \cdot {\log} \big|\mathbf I + \mathbf H_k \mathbf F_k \mathbf F_k^{\textsf H} \mathbf H_k^{\textsf H} \mathbf Q_k^{-1}\big|, ~\forall k \in {\cal N},
\end{equation}
where $\mathbf Q_k = \epsilon^2 \mathbf I + \sum\nolimits_{i = k+1}^{N} \mathbf H_i \mathbf F_i \mathbf F_i^{\textsf H} \mathbf H_i^{\textsf H}$, $\forall k \in {\cal N}_{-1}$, and $\mathbf Q_N = \epsilon^2 \mathbf I$. In the following, we define $\mathbf S_k \triangleq \mathbf F_k \mathbf F_k^{\textsf H}$ for convenience, $\forall k \in {\cal N}$.

Let $\beta_{k} \in [0,1]$ denote the offloading ratio of the $k$-th MU, and the uplink offloading time can then be expressed as
\begin{equation}
    T^{off}_k = \frac{\beta_k L_k}{R_k}, ~\forall k \in {\cal N}.
\end{equation}
The corresponding transmit energy consumption of the $k$-th MU is given by
\begin{equation}
E_k^{off} = T_k^{off} \text{Tr}(\mathbf S_k), ~\forall k \in {\cal N}.
\end{equation}
According to \cite{Xidian}, the local execution time of the $k$-th MU can be written as
\begin{equation}
	T_k^{local} = \frac{(1 - \beta_k) C_k L_k}{f_k},~\forall k \in {\cal N},
\end{equation}
where $f_{k}$ denotes the computational frequency of the $k$-th MU and it can be dynamically adjusted according to the task load. Following \cite{Xidian}, we denote the computational energy consumption of the $k$-th MU as
\begin{equation}
	E_k^{local} = \eta (1 - {\beta_k}){C_k}{L_k}f_k^2, ~\forall k \in {\cal N},
\end{equation}
where $\eta$ is the CPU's computation coefficient depending on chip architecture. Note that the total execution time of an offloaded task consists of the uplink and downlink transmission time, the execution time at the MEC server, and the local execution time at MUs. Regarding the transmission time, it is worth noting that we only consider the uplink transmission time and ignore the downlink transmission time. The reason is explained as follows: compared with the data size of the original task, the size of the computation result is negligible, and therefore the time of sending it from the MEC server to the MU can be ignored \cite{onlyuplink}. Moreover, we follow \cite{multinoma} and ignore the task execution time at the MEC server since the computational resource therein can be regarded as sufficient.

We aim to minimize the total energy consumption of all MUs. To this end, we formulate the following optimization problem:
\begin{subequations}
    \begin{eqnarray}
        \hspace{-0.75cm} \text{P1}: & \min\limits_{\bm{f}, \bm {\beta}, {\cal S}} & \sum\limits_{k = 1}^N \left(E_k^{off} + E_k^{local} \right) \\[1ex]
	    \hspace{-0.75cm} & \text{s.t.} &  0 \le f_k \le f_{\max}, \forall k \in {\cal N}, \\[1ex]
	    \hspace{-0.75cm} &&  0 \le \beta _k \le 1, \forall k \in {\cal N}, \\[1ex]
	    \hspace{-0.75cm} &&  0 \le \text{Tr}(\mathbf S_k) \le P_{\max}, \forall k \in {\cal N}, \\[1ex]
	    \hspace{-0.75cm} &&  T_k^{off} \le T_k, \forall k \in {\cal N}, \\[1ex]
	    \hspace{-0.75cm} &&  T_k^{local} \le T_k, \forall k \in {\cal N},
    \end{eqnarray}
\end{subequations}
where $\bm f = [f_1, \cdots, f_N]$, $\bm \beta = [\beta_1, \cdots, \beta_N]$, and ${\cal S} = \left\{\mathbf S_1, \cdots, \mathbf S_N\right\}$. (7b), (7c), and (7d) respectively denote the constraints of maximum computational resource, offloading ratio, and transmit power. Moreover, since local computing and remote offloading can be executed simultaneously \cite{mumimo}, (7e) and (7f) jointly characterize the constraint of maximum processing delay. Due to (7a) and (7e), P1 is a non-convex optimization problem which cannot be solved directly.

\section{Computation and Communication Co-Design}
In this section, we decompose P1 into three subproblems and solve them iteratively until convergence, as detailed below.

\subsection{Computational Resource Optimization}
Firstly, we fix $\bm {\beta}$ and $\cal S$ and consider the optimization of $\bm f$. The corresponding subproblem is formulated as
\begin{subequations}
    \begin{eqnarray}
    \text{P2}: & \min\limits_{\bm f} & \sum\limits_{k = 1}^N {\eta (1 - \beta_k){C_k}{L_k} f_k^2} \\[1ex]
    & \text{s.t.} & 0 \le f_k \le f_{\max}, \forall k \in {\cal N}, \\[2ex]
    & & \frac{(1 - \beta_k) C_k L_k}{f_k} \le T_k, \forall k \in {\cal N}.
    \end{eqnarray}
\end{subequations}
Since the objective function in P2 is monotonically increasing with respect to each $f_k$, the minimum objective value is obtained when 
\begin{equation} \label{Optimal-f}
    f_k = \frac{(1 - \beta _k) C_k L_k} {T_k}    
\end{equation}
holds, which establishes a unique mapping between $f_k^{\star}$ and $\beta_k^{\star}$, $\forall k \in {\cal N}$. By substituting \eqref{Optimal-f} into P1 and removing $\bm f$ and its related constraint, P1 is simplified as follows
\begin{subequations}
\begin{align}
    \text{P3}: \min\limits_{\bm \beta, {\cal S}} \enspace & \sum\limits_{k = 1}^N \left[\frac{\beta_k L_k \text{Tr}(\mathbf S_k)}{R_k} + \frac{\eta C_k^3 L_k^3 (1 - \beta_k)^3}{T_k^2} \right] \\[2ex]
    \text{s.t.} \enspace \enspace  & 0 \le \beta _k \le 1, \forall k \in {\cal N}, \\[1ex]
    & \frac{\beta_k L_k}{R_k} \le {T_k},\forall k \in {\cal N}, \label{P3-DelayConstraint} \\[1ex]
    & \text{Tr}(\mathbf S_k) \le P_{\max}, \forall k \in {\cal N}
\end{align}
\end{subequations}

\subsection{Offloading Ratio Optimization}
We then fix $\cal S$ and focus on the optimization of $\bm \beta$. The corresponding subproblem is formulated as
 \begin{subequations}
 \begin{align}
    \text{P4}: \min\limits_{\bm \beta} \enspace & \sum\limits_{k = 1}^N \left[\frac{\beta_k L_k \text{Tr}(\mathbf S_k)}{R_k} + \frac{\eta C_k^3 L_k^3 (1 - \beta_k)^3}{T_k^2} \right] \\[2ex]
    \text{s.t.} \enspace \enspace & 0 \le \beta_k \le 1, \forall k \in {\cal N}, \label{P4-betaConstraint} \\[2ex]
    & \beta_k \le \frac{R_k T_k}{L_k}, \forall k \in {\cal N}, \label{P4-DelayConstraint}
 \end{align}
 \end{subequations}
where we have rewritten \eqref{P3-DelayConstraint} in the form of \eqref{P4-DelayConstraint}. Note that $\bm \beta = [\beta_1, \cdots, \beta_N]$ in P4 can be decoupled, and therefore we can optimize each $\beta_k$ individually employing the following lemma.
\begin{lemma}\label{Convexity-Monotonicity-Lemma}
    $\Omega(\beta_k) = a_k (1 - \beta_k)^3 + b_k \beta_k$, where $a_k = \eta \frac{C_k^3 L_k^3}{T_k^2}$ and $b_k = \frac{\text{Tr}(\mathbf S_k) L_k}{R_k}$, is a convex function between 0 and 1, and it has a stationary point $\beta_k^{+} = 1 - \sqrt{\frac{b_k}{3a_k}}$.
\end{lemma}
\begin{IEEEproof}
    To find the stationary point of $\Omega(\beta_k)$, we let ${\partial \Omega(\beta_k)} / {\partial \beta_k} = 0$, and obtain that $\beta_k^{+} = 1 - \sqrt{b_k / (3 a_k)}$; the other solution of ${\partial \Omega(\beta_k)} / {\partial \beta_k} = 0$, i.e., $1 + \sqrt{b_k / (3 a_k)}$, is definitely out of the feasible region of $\beta_k$, which is $[0, \min({R_k T_k}/{L_k}, 1)]$ according to \eqref{P4-betaConstraint} and \eqref{P4-DelayConstraint}. As for the convexity, we compute the second order derivative of $\Omega(\beta_k)$ with respect to $\beta_k$, given by ${\partial \Omega^2(\beta_k)} / {\partial \beta^2_k} = 6 a_k (1 - \beta _k)$. It is observed that ${\partial \Omega^2(\beta_k)} / {\partial \beta^2_k} \ge 0$, $\forall \beta_k \in [0, 1]$, which means that $\Omega(\beta_k)$ is a convex function. Until now, Lemma \ref{Convexity-Monotonicity-Lemma} has been proven.
\end{IEEEproof}

According to Lemma \ref{Convexity-Monotonicity-Lemma} and the relationships among 0, 1, $\beta_k^{+}$, and $R_k T_k / L_k$, we derive the optimal $\beta_k^\star$ considering the following three cases.
\begin{itemize}
    \item $0 < \beta_k^{+} \le \min(R_k T_k / L_k, 1)$. In this case, $\Omega(\beta_k)$ monotonically decreases in the range of $[0, \beta_k^{+}]$ and increases in the range of $[\beta_k^{+}, \min(R_k T_k / L_k, 1)]$. Thus, $\beta_k^{\star} = \beta_k^{+} = 1 - \sqrt{b_k / (3a_k)}$, $\forall k \in {\cal N}$.
    \item $R_k T_k / L_k \le \beta_k^{+} \le 1$. In this case, $\beta_k \in [0, R_k T_k / L_k]$, and $\Omega(\beta_k)$ monotonically decreases in this range, such that $\beta_k^{\star} = R_k T_k / L_k$, $\forall k \in {\cal N}$.
    \item $\beta_k^{+} \le 0$. In this case, $\Omega(\beta_k)$ is a monotonically increasing function of $\beta_k$ in the range of $[0, \min(R_k T_k / L_k, 1)]$, and hence the optimal $\beta_k^{\star}$ that minimizes $\Omega(\beta_k)$ is 0, i.e., $\beta_k^{\star} = 0$, $\forall k \in {\cal N}$.
\end{itemize}

\subsection{Transmit Precoding Matrix Optimization}
We then fix $\bm \beta$ and consider the optimization of ${\cal S}$. The corresponding subproblem is formulated as
\begin{subequations}
\begin{align} \label{P5a}
    \text{P5}: \min\limits_{\cal S} \enspace & \sum\limits_{k = 1}^N \frac{\beta_k L_k \text{Tr}(\mathbf S_k)}{B \cdot \log \left|\mathbf I + \mathbf H_k \mathbf S_k \mathbf H_k^{\textsf H} \mathbf Q_k^{-1}\right|} \\[2ex]
    \text{s.t.} \enspace \enspace & \text{Tr}(\mathbf S_k) \le P_{\max}, \forall k \in {\cal N}, \label{P5b} \\[1ex]
    & {\log} \left|\mathbf I + \mathbf H_k \mathbf S_k \mathbf H_k^{\textsf H} \mathbf Q_k^{-1}\right| \ge \frac{\beta_k L_k}{T_k B}, \forall k \in {\cal N}. \label{P5c}
\end{align}
\end{subequations}
It can be observed that P5 is still challenging to solve due to the non-convexity of \eqref{P5a} and \eqref{P5c}. In the following, we adopt the alternative optimization (AO) strategy and further decompose P5 into $N$ subproblems, where the first subproblem is given by
\begin{align*}
    \text{P5-1}: \min\limits_{\mathbf S_1} \enspace & \frac{\beta_1 L_1 \text{Tr}(\mathbf S_1)} {B \cdot \log\left|\mathbf I + \mathbf H_1 \mathbf S_1 \mathbf H_1^{\textsf H} \mathbf Q_1^{-1}\right|} + \sum\limits_{k = 2}^N E_k^{off}  \\[2ex]
    \text{s.t.} \enspace \enspace & \text{Tr}(\mathbf S_1) \le P_{\max}, \\[1ex]
    & \log \big|\mathbf I + \mathbf H_1 \mathbf S_1 \mathbf H_1^{\textsf H} \mathbf Q_1^{-1}\big| \ge \frac{\beta_1 L_1}{T_1 B},
\end{align*}
and the $j$-th subproblem, $\forall j \in \{2, \cdots, N\}$, is given by
\begin{align*}
    \text{P5-2}: \min\limits_{\mathbf S_j} \enspace & \sum\limits_{k = 1}^{j-1} \frac{\beta_k L_k \text{Tr}(\mathbf S_k)}{B \cdot \log\left|\mathbf I + \mathbf H_k \mathbf S_k \mathbf H_k^{\textsf H} [\mathbf Q_k(\mathbf S_j)]^{-1}\right|} \nonumber \\[1ex]
    & + \frac{\beta_j L_j \text{Tr}(\mathbf S_j)}{B \cdot \log \left|\mathbf I + \mathbf H_j \mathbf S_j \mathbf H_j^{\textsf H} \mathbf Q_j^{-1}\right|} + \sum\limits_{k = j + 1}^N E_k^{off} \\[2ex]
	\text{s.t.} \enspace & \text{Tr}(\mathbf S_j) \le P_{\max}, \\[1ex]
	& \log\big|\mathbf I + \mathbf H_j \mathbf S_j \mathbf H_j^{\textsf H} \mathbf Q_j^{-1}\big| \ge \frac{\beta_j L_j}{T_j B}, \\
	& \log \big|\mathbf I + \mathbf H_k \mathbf S_k \mathbf H_k^{\textsf H} \mathbf Q_k^{-1}\big| \ge \frac{\beta_k L_k}{T_k B}, \forall k \in {\cal J}_{-1},
\end{align*}
with ${\cal J}_{-1} \triangleq \{1, \cdots, j - 1\}$.

Below we introduce how to efficiently solve P5-1 and P5-2. To solve P5-1, we first decompose $\mathbf Q_1^{-1} = \big(\epsilon^2 \mathbf I + \sum\nolimits_{k = 2}^N \mathbf H_k \mathbf S_k \mathbf H_k^{\textsf H}\big)^{-1}$ as $\mathbf Q_1^{-1} = \mathbf A_1^{\textsf H} \mathbf A_1$, and then rewrite $\log \left|\mathbf I + \mathbf H_1 \mathbf S_1 \mathbf H_1^{\textsf H} \mathbf Q_1^{-1}\right|$ as
\begin{equation}
    {\log} \left|\mathbf I + \mathbf H_1 \mathbf S_1 \mathbf H_1^{\textsf H} \mathbf Q_1^{-1}\right| = {\log} \left|\mathbf I + \mathbf A_1 \mathbf H_1 \mathbf S_1 \mathbf H_1^{\textsf H} \mathbf A_1^{\textsf H} \right|.
\end{equation}
Subsequently, we do singular value decomposition for $\mathbf A_1 \mathbf H_1$, denoted by $\mathbf A_1 \mathbf H_1 = \mathbf U_1 \bm \Sigma_1 \mathbf V_1^{\textsf H}$. We let $\mathbf S_1 = \mathbf V_1 \bm \Lambda_1 \mathbf V_1^{\textsf H}$ and rewrite ${\log} \left|\mathbf I + \mathbf A_1 \mathbf H_1 \mathbf S_1 \mathbf H_1^{\textsf H} \mathbf A_1^{\textsf H} \right|$ as
\begin{equation}
    \log \left|\mathbf I + \mathbf A_1 \mathbf H_1 \mathbf S_1 \mathbf H_1^{\textsf H} \mathbf A_1^{\textsf H} \right| = \sum\limits_{i=1}^{N_t} \log(1 + \lambda_{1,i} \sigma^2_{1,i}),
\end{equation}
where $\lambda_{1,i}$ and $\sigma_{1,i}$ respectively denote the $i$-th diagonal entry of $\bm \Lambda_1$ and $\bm \Sigma_1$. Moreover, since $\text{Tr}(\mathbf S_1) = \text{Tr}(\bm \Lambda_1)$, we reformulate P5-1 as follows
\begin{subequations}
\begin{eqnarray}
    \text{P6-1}: & \min\limits_{\bm \lambda_1} & \frac{\beta_1 L_1 \sum\limits_{i=1}^{N_t} \lambda_{1, i}}{B \cdot \sum\limits_{i=1}^{N_t} \log\left(1 + \lambda_{1,i} \sigma^2_{1,i}\right)} \\[1ex]
    & \text{s.t.} & \sum\limits_{i=1}^{N_t} \lambda_{1, i} \le P_{\max}, \label{P61-PowerCons} \\
    & & \sum\limits_{i=1}^{N_t} \log(1 + \lambda_{1,i} \sigma^2_{1,i}) \ge \frac{\beta_1 L_1}{T_1 B}, \label{P61-RateCons}
\end{eqnarray}
\end{subequations}
where $\bm \lambda_1 = [\lambda_{1, 1}, \cdots, \lambda_{1, N_t}]$. To solve P6-1, we introduce an auxiliary variable $\delta_1 > 0$ and transform it as
\begin{subequations}
\begin{eqnarray}
    \hspace{-0.75cm} \text{P7-1}: & \min\limits_{\bm \lambda_1} & \delta_1 \\
    \hspace{-0.75cm} & \text{s.t.} & \frac{\beta_1 L_1}{\delta_1 B} \sum\limits_{i=1}^{N_t} \lambda_{1, i} \le \sum\limits_{i=1}^{N_t} \log\left(1 + \lambda_{1,i} \sigma^2_{1,i}\right), \label{P71-QuasiConvexCons} \\[2ex] 
    \hspace{-0.75cm} & & \eqref{P61-PowerCons}, ~\eqref{P61-RateCons}.
\end{eqnarray}
\end{subequations}
Then, we leverage the bisection method to solve P7-1. Specifically, for a fixed value of $\delta_1$, we construct the following feasibility problem
\begin{subequations}
\begin{align}
\text{P8-1}: \text{find} \enspace & \bm \lambda_1 \\[1ex]
\text{s.t.}  \enspace & \eqref{P71-QuasiConvexCons}, ~\eqref{P61-PowerCons}, ~\eqref{P61-RateCons}.
\end{align}
\end{subequations}
If this problem is feasible, then the optimal objective value of P6-1 is less than or equal to $\delta_1$. While if P8-1 is infeasible, then the optimal objective value of P6-1 is larger than $\delta_1$.

To solve P5-2, we introduce $t_1, \cdots, t_{j - 1} > 0$ and $\delta_j > 0$, and transform it as follows.
\begin{subequations}\label{P62}
\begin{align}
	\hspace{-0.25cm} \min\limits_{\mathbf S_j, \delta_j, \{t_k\}} \enspace & \delta_j + \sum\limits_{k = 1}^{j-1} t_k \label{P62a} \\[1ex]
	\hspace{-0.25cm} \text{s.t.} \enspace & \frac{\beta_j L_j \text{Tr}(\mathbf S_j)}{\delta_j B} \le \log \big|\mathbf I + \mathbf H_j \mathbf S_j \mathbf H_j^{\textsf H} \mathbf Q_j^{-1}\big|, \label{P62b} \\[1ex]
	\hspace{-0.25cm} & \text{Tr}(\mathbf S_j) \le P_{\max}, \label{P62c} \\[2ex]
	\hspace{-0.25cm} & \frac{\beta_k L_k \text{Tr}(\mathbf S_k)}{t_k B} \le g_{k,1} - g_{k,2}, \forall k \in {\cal J}_{-1}, \label{P62d} \\[1ex]
	\hspace{-0.25cm} & \frac{\beta_j L_j}{T_j B} \le \log \big|\mathbf I + \mathbf H_j \mathbf S_j \mathbf H_j^{\textsf H} \mathbf Q_j^{-1}\big|, \label{P62e} \\[1ex]
	\hspace{-0.25cm} & \frac{\beta_k L_k}{T_k B} \le g_{k,1} - g_{k,2}, \forall k \in {\cal J}_{-1}, \label{P62f}
\end{align}
\end{subequations}
where $g_{k,1}(\mathbf S_j) = \log\big|\mathbf H_k \mathbf S_k \mathbf H_k^{\textsf H} + \epsilon^2 \mathbf I + \sum\nolimits_{i = k+1}^N \mathbf H_i \mathbf S_i \mathbf H_i^{\textsf H}\big|$ and $g_{k,2}(\mathbf S_j) = \log\big|\epsilon^2 \mathbf I + \sum\nolimits_{i = k+1}^N \mathbf H_i \mathbf S_i \mathbf H_i^{\textsf H}\big|$, $\forall k \in {\cal J}_{-1}$. Note that in \eqref{P62}, we have used the following property that
\begin{align*}
	\frac{R_k}{B} & =  \log \big|\mathbf I + \mathbf H_k \mathbf S_k \mathbf H_k^{\textsf H} (\epsilon^2 \mathbf I + \sum\limits_{i = k + 1}^{N} \mathbf H_i \mathbf S_i \mathbf H_i^{\textsf H})^{-1}\big| \\
	& = g_{k,1} - g_{k,2}.
\end{align*}
We can observe that \eqref{P62} is still a non-convex optimization problem due to \eqref{P62b}, \eqref{P62d} and \eqref{P62f}. To tackle \eqref{P62b}, we introduce a new variable $\xi > 0$, and let $\text{Tr}(\mathbf S_j) = \xi^2$. Then, we can rewrite \eqref{P62b} as
\begin{equation}\label{P62b-Transformation}
    \frac{\beta_j L_j \xi^2}{\delta_j B} \le \log \left|\mathbf I + \mathbf H_j \mathbf S_j \mathbf H_j^{\textsf H} \mathbf Q_j^{-1}\right|.
\end{equation}
To cope with the non-convexity of $\text{Tr}(\mathbf S_j) = \xi^2$, we introduce a penalty parameter $\Delta > 0$ and rewrite \eqref{P62} as follows
\begin{subequations} \label{P72}
\begin{align}
    \hspace{-1cm} \min\limits_{\mathbf S_j, \delta_j, \xi, \{t_k\}} & \delta_j + \Delta [\text{Tr}(\mathbf S_j) - \xi^2] + \sum\limits_{k = 1}^{j-1} t_k \label{P72a} \\[1ex]
    \hspace{-1cm} \text{s.t.} \enspace & \text{Tr}(\mathbf S_j) \ge \xi^2, \label{P72b} \\[2ex]
    \hspace{-1cm} & \eqref{P62c} - \eqref{P62f}, ~\eqref{P62b-Transformation}.
\end{align}
\end{subequations}
Then, we use the successive convex approximation (SCA) technique to tackle the non-convexity of \eqref{P72a}, \eqref{P62d} and \eqref{P62f}. Specifically, at the $n$-th SCA iteration, by linearizing $\xi^2$ and $g_{k,2}$, i.e., $\xi^2 \ge (\xi^{[n]})^2 + 2 \xi^{[n]} (\xi - \xi^{[n]})$ and
\begin{align*}
	& g_{k,2} = \log \big|\mathbf H_j \mathbf S_j \mathbf H_j^{\textsf H} + \mathbf Q_{k,-j} \big| \le \log \big|\mathbf H_j \mathbf S^{[n]}_j \mathbf H_j^{\textsf H} + \mathbf Q_{k,-j} \big| \\[2ex]
	& ~~+ \text{Real}\big\{\text{Tr}\big[\big(\mathbf H_j \mathbf S^{[n]}_j \mathbf H_j^{\textsf H} + \mathbf Q_{k,-j}\big)^{-1} \mathbf H_j (\mathbf S_j - \mathbf S^{[n]}_j) \mathbf H_j^{\textsf H}\big]\big\} \\[2ex]
	& ~~\triangleq {\tilde g}_{k,2},
\end{align*}
where $\mathbf Q_{k,-j} = \epsilon^2 \mathbf I + \sum\limits_{i = k + 1, i \ne j}^{N} \mathbf H_i \mathbf S_i \mathbf H_i^{\textsf H}$, we successfully construct a convex optimization problem, given by
\begin{subequations} \label{P82}
\begin{align}
    \hspace{-0.2cm} \min\limits_{\mathbf S_j, \delta_j, \xi, \{t_k\}} \enspace & \delta_j + \Delta (\text{Tr}(\mathbf S_j) + (\xi^{[n]})^2 - 2 \xi^{[n]} \xi) + \sum\limits_{k = 1}^{j-1} t_k \\[1ex]
    \hspace{-0.2cm} \text{s.t.} \enspace & \frac{\beta_k L_k \text{Tr}(\mathbf S_k)}{t_k B} \le g_{k,1} - {\tilde g}_{k,2}, ~\forall k \in {\cal J}_{-1}, \\[1ex]
    \hspace{-0.2cm} & \frac{\beta_k L_k}{T_k B} \le g_{k,1} - {\tilde g}_{k,2}, ~\forall k \in {\cal J}_{-1}, \\[2ex]
    \hspace{-0.2cm} & \eqref{P62c}, ~\eqref{P62e}, ~\eqref{P62b-Transformation}, ~\eqref{P72b}.
\end{align}
\end{subequations}
Up to now, we have introduced how to solve P5-2 efficiently. Then, by solving P5-1 and P5-2 iteratively until convergence, we obtain a locally optimal solution to P5. Last, by iteratively solving P4 and P5, a locally optimal solution to P3 is obtained.

\section{Simulation Results}
In this section, we evaluate the performance of the proposed algorithm by simulations. It is assumed that the considered MIMO-NOMA enabled MEC system consists of one BS and $N = 2$ MUs, and the channel between MU and BS follows Rayleigh fading with zero mean and variance $10^{-5}$ (large scaling fading). The other parameters used in this paper are summarized as follows: $P_{\max} = 1$ W, $N_0 = -174$ dBm/Hz, $B = 25$ MHz, $L_1 = L_2 = 5$ MB, $C_1 = C_2 = 200$ cycles/bit, $f_1 = f_2 = 2\text{G}$ cycles/second, and $\eta = 10^{-32}$.

\begin{figure}[t]
	\centering
	\includegraphics[scale=0.45]{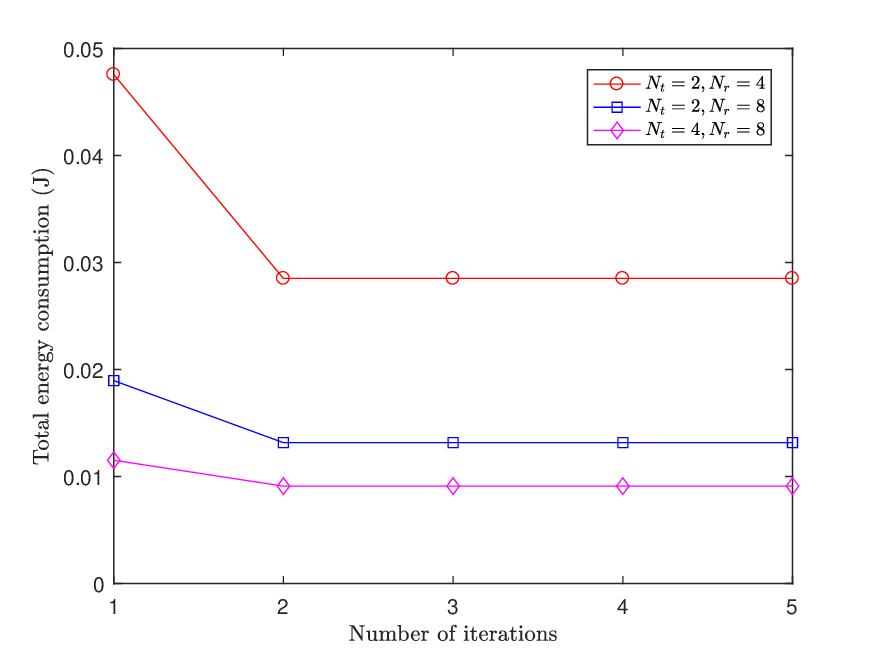}
	\caption{Total energy consumption with respect to the number of iterations. $T_1 = T_2 = 0.5$ second.} \label{Fig:2}
	\vspace{-1em}
\end{figure}

Fig. \ref{Fig:2} depicts the convergence behaviour of the proposed algorithm. From this figure, we immediately observe that the total energy consumption of the two MUs decreases after each iteration and the proposed algorithm converges quickly in only a few iterations. Moreover, we also observe that the total energy consumption decreases when we increase the number of transmit/receive antennas. The reasons can be explained as follows. When we increase the number of receive antennas, a higher signal-to-noise ratio can be achieved at the BS, such that the transmission rates of the two MUs increase, and accordingly both the offloading time and the total energy consumption decrease. When we increase the number of transmit antennas, the offloading time and the total energy consumption also decrease since a higher degree-of-freedom is achieved in this case.

\begin{figure}[t]
	\centering
	\includegraphics[scale=0.45]{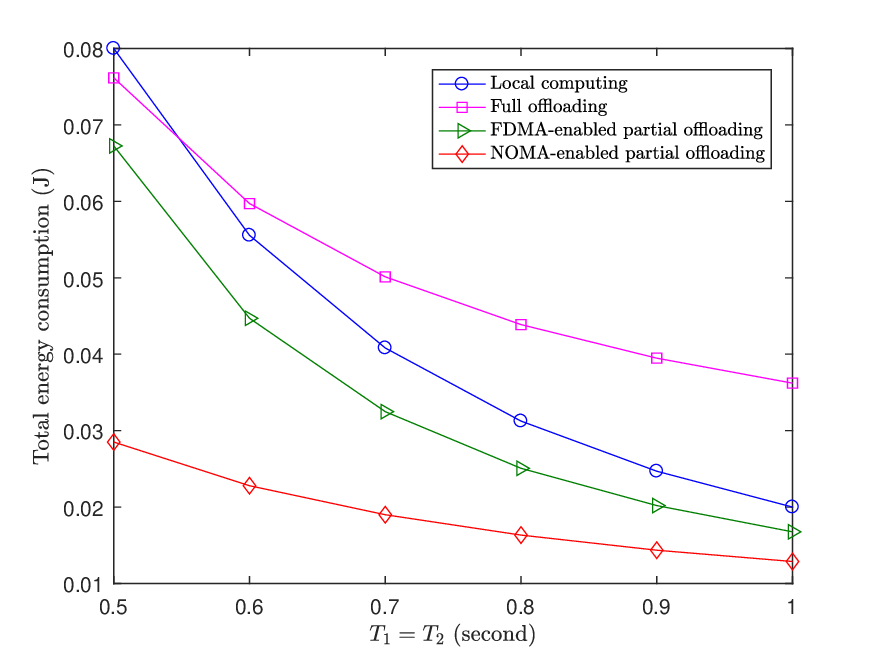}
	\caption{Total energy consumption with respect to the task tolerable latency. $N_t = 2$, and $N_r = 4$.} \label{Fig:3}
	\vspace{-1em}
\end{figure}

Fig. \ref{Fig:3} shows the effect of task tolerance latency on the total energy consumption of the two MUs. In addition to the proposed algorithm, the following three methods are also considered for comparison: 1) Local computing, where both the two MUs computer their respective tasks locally, i.e., $\beta_1 = \beta_2 = 0$; 2) Full offloading, i.e., $\beta_1 = \beta_2 = 1$; and 3) Frequency division multiple access (FDMA)-enabled partial offloading \cite{mumimo, decoding2}. As expected, with the increase of maximum tolerance latency, the total energy consumption of the two MUs for the four methods all decrease. We also observe that our proposed MIMO-NOMA enabled partial offloading scheme consumes the least energy.


\section{Conclusion}
In this letter, we aimed to minimize the total energy consumption of MUs for MIMO-NOMA enabled MEC, where a joint offloading ratio and transmit precoding optimization algorithm was proposed. For tractability, we decomposed the originally formulated problem into three subproblems, which were solved iteratively until convergence. Simulation results verified that the application of MIMO-NOMA to MEC indeed reduces the total energy consumption of MUs compared to the FDMA-enabled MEC, and partial offloading indeed outperforms full offloading and local computing.

\end{document}